# Biomechanical and Mechanobiological Modelling of Functionally Graded Scaffolds for Large Bone Defects

Ali Entezari [1,2] *[‡], Vahid Badali [2,3][‡], Sara Checa [2,4] *

[1]School of Biomedical Engineering, Faculty of Engineering and Information Technology, University of Technology Sydney, Sydney, NSW 2007, Australia
[2]Julius Wolff Institute, Berlin Institute of Health, Charité—Universitätsmedizin Berlin, Berlin, Germany
[3]Department of Structural Mechanics and Analysis, Technische Universität Berlin, Berlin, Germany
[4]Institute of Biomechanics, TUHH Hamburg University of Technology, Hamburg, Germany

* Corresponding authors: ali.entezari@uts.edu.au and sara.checa@tuhh.de
[‡] Ali Entezari and Vahid Badali contributed equally to this work.

**Abstract**

Critical-sized bone defects remain a major clinical challenge, requiring scaffolds that combine mechanical stability with regenerative capacity. Functionally graded (FG) scaffolds, inspired by the graded architecture of native bone, offer a promising solution by spatially varying porosity to optimise both load transfer and tissue ingrowth. Here, we present an integrated finite element–agent-based modelling (FEA–ABM) framework to simultaneously evaluate the biomechanics and regenerative potential of FG scaffolds under physiologically relevant conditions. Cylindrical scaffolds with axial or radial pore-size gradients were compared with uniform controls. The finite element model incorporated poroelastic tissue mechanics and gait-related loading to compute local shear strain and fluid velocity, which guided cellular behaviours in the agent-based model, including progenitor migration, proliferation, differentiation, and apoptosis. Simulations over 150 days revealed that axial gradients with larger pores at the host bone interface promoted greater bone ingrowth, while radial gradients with denser peripheral struts substantially reduced peak von Mises stresses. These findings highlight a fundamental design trade-off between maximising regenerative performance and enhancing structural competence. The coupled FEA–ABM framework establishes a mechanistic platform for the rational design of next-generation FG scaffolds, offering a pathway toward pre-clinical optimisation of implants tailored to defect location and loading environment.

# 1. Introduction

Treatment of bone defects resulting from trauma, tumour resection, or degenerative disease remains a major challenge in orthopaedic surgery [1]. Current clinical options for reconstruction primarily rely on autografts and allografts, but both approaches present serious limitations such as limited availability, donor-site morbidity, risk of immune rejection, and disease transmission [2, 3]. A promising alternative is bone tissue engineering, which seeks to regenerate bone by combining biological factors with engineered constructs [4]. Central to this approach are porous three-dimensional scaffolds, typically fabricated from biocompatible and osteoconductive materials, which not only provide mechanical stability but also act as a substrate to guide cellular infiltration, vascularisation, and new bone formation within the defect site [5].

Despite considerable progress in bone tissue engineering over recent decades, the treatment of critical-sized defects remains far from optimal [6]. A central challenge lies in achieving the right balance between mechanical strength and regenerative capacity [5]. Scaffolds must not only provide sufficient stability to withstand physiological loads in weight-bearing sites, but also facilitate effective bone regeneration [7]. To address this dual requirement, functionally graded (FG) scaffolds have emerged as a promising strategy [8]. These scaffolds are inspired by natural bone, which displays gradual architectural transitions across regions of differing density and porosity. Mimicking such gradients, recent studies have explored FG scaffold designs in which architecture is varied spatially to enhance both mechanical performance and tissue regeneration [9, 10].

Recent studies support FG scaffold designs as a means to reconcile mechanical competence with osteogenesis [11]. Pore-size gradients, implemented axially or radially, have been shown to tune stiffness, stress distribution, and fluid transport properties, thereby improving cell attachment and overall regenerative potential compared with uniform architectures [12]. For example, Xu et al. demonstrated that gradient pore-change strategies in triply periodic minimal surface scaffolds fabricated via selective laser sintering result in distinct mechanical responses and fluid-flow characteristics [13]. Similarly, Di Luca et al. showed that introducing pore-size gradients enhances the osteogenic differentiation of human mesenchymal stromal cells, highlighting the biological relevance of graded architectures for bone regeneration [14]. More recently, Xiao et al. reported that gradient Schwarz Primitive scaffolds with thin-board integration achieved improved mechanical stability and promoted osteoblast adhesion and proliferation, further underscoring the promise of graded architectures for clinical translation [15]. While these studies have provided valuable insights into the role of FG scaffolds in osteogenesis, most remain confined to in vitro settings and do not capture the complex mechano-biological environment of specific anatomical sites, which is critical for guiding scaffold optimisation.

Computational approaches have shown great potential as tools to evaluate the biomechanical properties of scaffolds and to optimise their design prior to experimental validation [16, 17]. Finite element analysis (FEA) has been widely employed to predict stress distribution, effective stiffness, and load transfer in porous structures, while mechanobiological models have been used to simulate scaffold-guided bone regeneration [18, 19]. By integrating these methods, it becomes possible to systematically explore scaffold architectures and identify optimal designs that balance mechanical competence with biological performance [20]. For example, Rezapourian and Hussainova recently performed a finite element study on hydroxyapatite-based Voronoi scaffolds with graded design parameters, showing that applying gradients in pore seed spacing and strut thickness improved load-bearing capacity and resulted in more favourable stress distribution compared to uniform scaffolds, highlighting the potential of irregular graded lattices to better mimic the structure and function of bone [21]. More recently, Wu et al. introduced a machine learning–driven dynamic optimisation framework to design functionally graded ceramic scaffolds with vertical and lateral porosity gradients, reporting that both gradient strategies enhanced long-term bone regeneration compared to uniform designs, with lateral gradients producing the most pronounced improvements [22].

While computational approaches have shown great promise in evaluating and optimising FG scaffolds for bone regeneration, most have relied on mechanobiological tissue differentiation models, where local mechanical stimuli are translated into rules for tissue formation and maturation. These algorithms successfully link mechanical cues to tissue differentiation but generally overlook the critical role of cellular behaviors such as proliferation, migration, and apoptosis. To address this limitation, agent-based models (ABMs) have been developed, enabling explicit representation of individual cell activities and their interactions with the scaffold microenvironment, thereby providing a more mechanistic framework to study scaffold-guided bone regeneration [23-25]. For example, Perier-Metz et al. coupled an ABM with finite element analysis to simulate progenitor cell migration, proliferation, differentiation, and apoptosis within large bone defects, allowing prediction of how scaffold design parameters influence bone regeneration outcomes [26]. Similarly, Jaber et al. applied a hybrid ABM–FEA framework to capture mesenchymal cell migration, osteoblast proliferation and differentiation, and fibroblast activity within scaffold pores, enabling comparison of regeneration dynamics between gyroid and strut-like architectures in a large femoral defect [27]. Nevertheless, previous ABM–FEA frameworks have been restricted to uniform scaffolds; none have examined FG architectures under physiologically relevant loading.

In this study, we present for the first time an integrated FEA–ABM framework to evaluate both the mechanical performance and regenerative capacity of functionally graded (FG) scaffolds under physiologically relevant conditions. Specifically, we investigate two types of pore-size gradients, axial and radial, each implemented in two opposing directions. Importantly, the study builds on our previously validated FEA–ABM framework, which integrates scaffold mechanics with explicit simulation of cellular processes that govern bone regeneration [23]. By systematically comparing these gradient strategies, our work elucidates how scaffold architecture shapes both biomechanical behaviour and regeneration dynamics, thereby providing design principles for next-generation FG scaffolds optimised for large bone defect repair.

## 2. Materials and Methods

### 2.1. Scaffold designs

Scaffolds with either radial or axial pore-size gradients were generated to investigate the influence of graded architectures on both mechanical performance and tissue regeneration. All scaffold models were designed as cylinders with a diameter of 21 mm and a height of 30 mm, matching the dimensions of the defect region.

For the axial gradients (Figure 1), two types were created. In Type 1, pore size increased from the top and bottom surfaces toward the central region, providing larger pores in the scaffold core while maintaining tighter pore networks at the host–scaffold interfaces. In Type 2, pore size decreased toward the central region, representing an exploratory design in which the effect of a denser mid-height on mechanical stability could be evaluated. A uniform scaffold with constant pore size was again included as a reference. To ensure comparability, all three scaffolds in this category were generated with a porosity of approximately 45%. The uniform scaffold had a pore size of 1.6 mm, while the largest and smallest pore sizes in the gradient scaffolds were 2.55 and 0.65 mm, respectively.

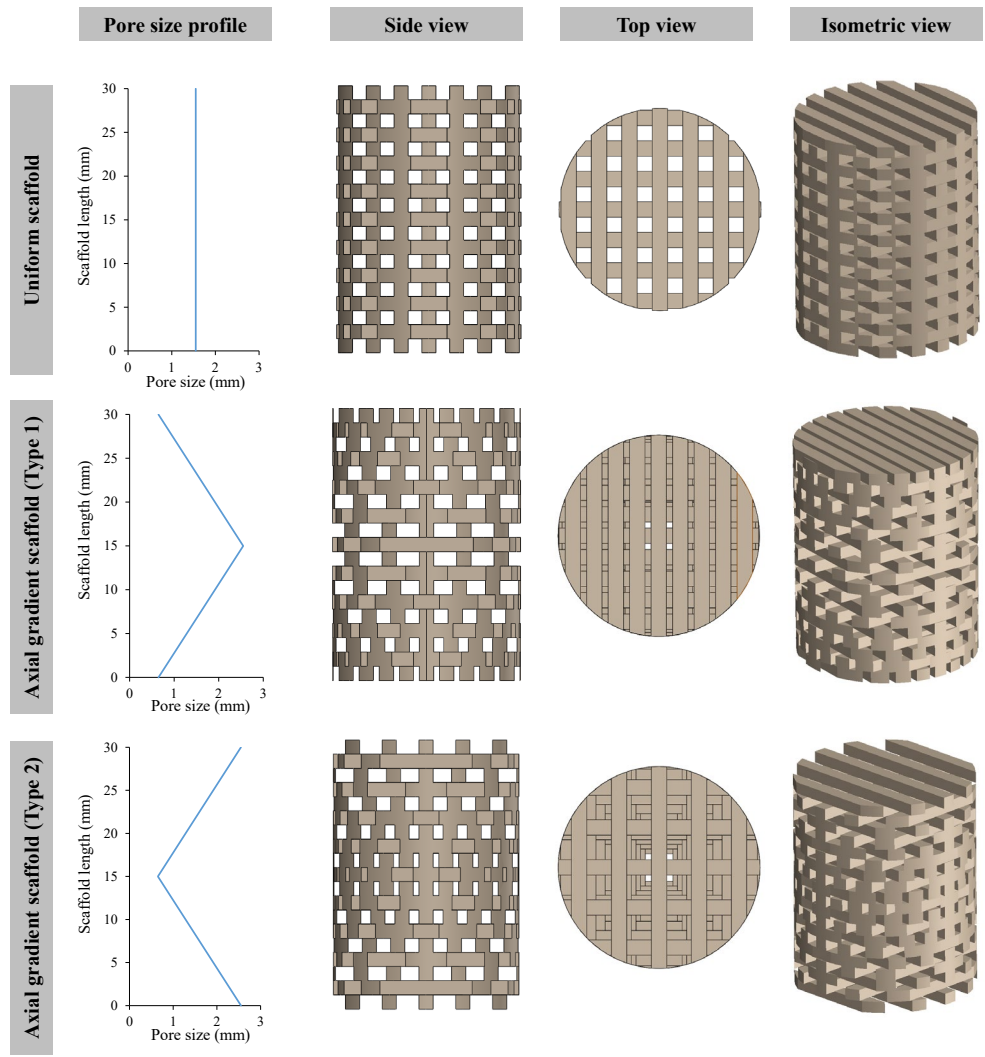

*Figure 1. Scaffold designs with different axial pore-size distributions.* The first row shows a uniform scaffold with constant pore size. The second row presents an axial gradient scaffold (Type 1), where pore size increases from the top and bottom surfaces toward the central region. The third row illustrates an axial gradient scaffold (Type 2), where pore size decreases from the surfaces toward the central region. For each scaffold type, the pore-size profile, side view, top view, and isometric view are provided to demonstrate architectural variations.

For the radial gradients (Figure 2), two types were created. In Type 1, pore size gradually decreased from the centre toward the periphery, whereas in Type 2 it decreased from the periphery toward the centre. The Type 1 design, with denser struts at the periphery, draws inspiration from the femoral architecture, where a stiff cortical shell surrounds a more porous trabecular core. A uniform scaffold with a constant pore size served as the control. As with the axial gradients, all three scaffolds in this category were designed with an overall porosity of approximately 45%. The uniform scaffold had a pore size of 1.5 mm. In Type 1, the largest and smallest pore sizes were

1.85 and 0.50 mm, respectively, whereas in Type 2 they were 2.05 and 0.55 mm. Notably, unlike the axial gradients where Types 1 and 2 exhibited similar maximum and minimum pore sizes, the radial gradient scaffolds required different pore sizes between Types 1 and 2 in order to maintain porosity consistency. The lower bound was set by biological considerations, as experimental studies indicate that pore sizes of at least ~300 µm are required to support adequate vascularisation and nutrient supply [28, 29].

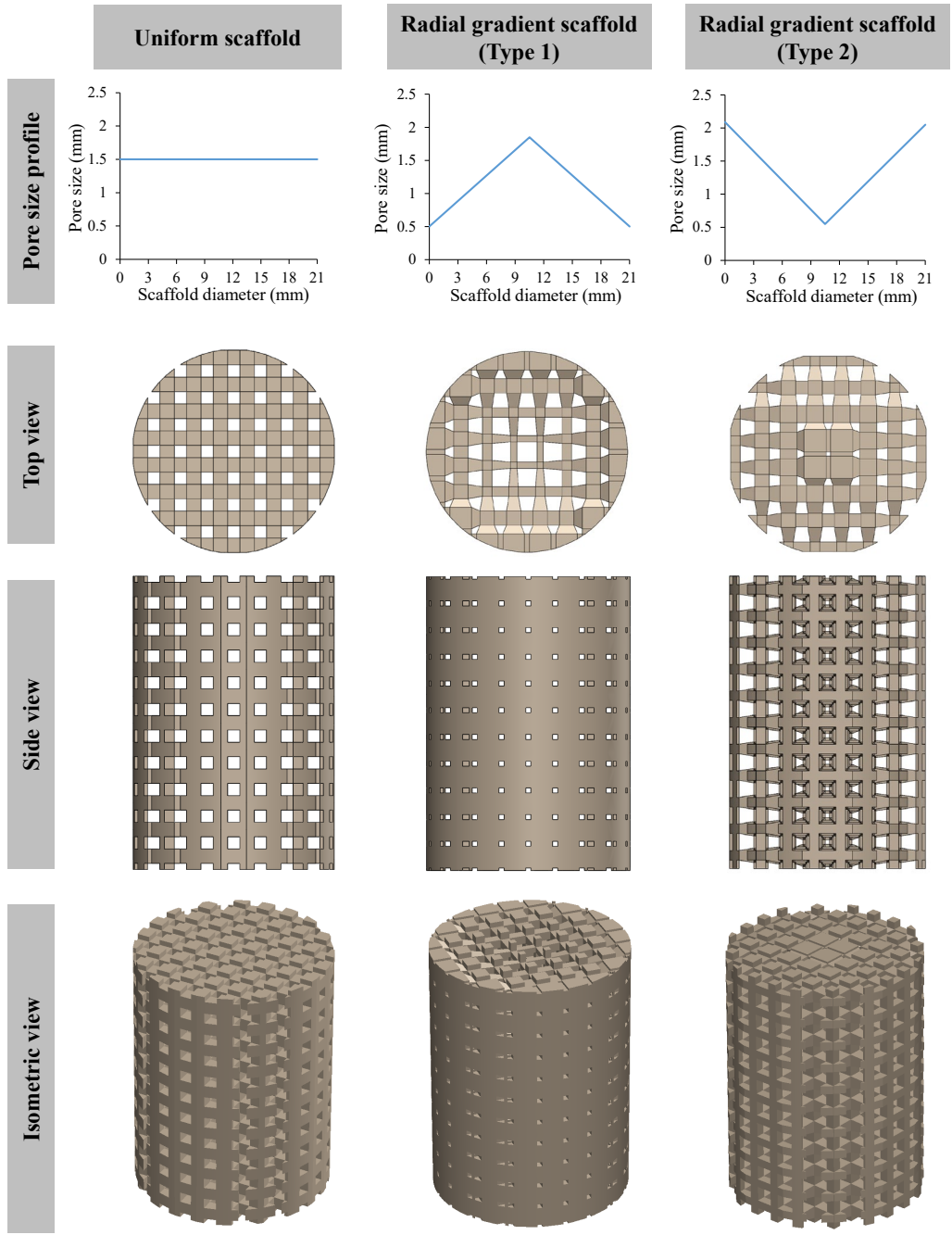

***Figure 1. Scaffold designs with different radial pore-size distributions.*** *The first column shows a uniform scaffold with constant pore size. The second column illustrates a radial gradient scaffold (Type 1), where pore size gradually increases from the periphery toward the centre. The third column depicts a radial gradient scaffold (Type 2), where pore size decreases from the periphery toward the centre. For each scaffold type, the pore-size profile, top view, side view, and isometric view are presented to highlight the architectural differences.*

**2.2 Integration of finite element and agent-based models**

**2.2.1. Finite element model**

A poroelastic finite element (FE) model was developed to characterize the mechanical environment within a large bone defect stabilized by a fixation plate and containing the scaffold (Figure 3a). Scaffold geometries were generated in SolidWorks (Dassault Systèmes, France) and imported into Abaqus/CAE (Dassault Systèmes, Providence, RI, USA), where they were positioned within an idealized cylindrical defect domain. The cortical bone, representing ovine femoral geometry, was modeled as a hollow cylinder with an outer diameter of 21 mm and an inner diameter of 15 mm. A callus domain was defined to surround the scaffold externally and to infiltrate the pore space within the scaffold, thereby enabling tissue deposition both around and inside the construct.

Biological tissues were represented as poroelastic materials with distinct elastic, permeability, and bulk modulus properties (Table 1), allowing the simultaneous consideration of solid and fluid phases throughout the healing environment [23, 24]. In contrast, the implant components were modeled as linear elastic solids: the scaffold was assigned the properties of Ti-6Al-4V alloy (Young's modulus 104 GPa, Poisson's ratio 0.3), and the fixation plate and screws were assigned the properties of 316L stainless steel (Young's modulus 200 GPa, Poisson's ratio 0.305) [30].

Physiological loading representative of ovine gait was applied. Specifically, an axial compressive load of 1,372 N—corresponding to approximately two times the body weight of a sheep—together with a bending moment of 17.125 Nm was imposed on the proximal bone extremity [31], while the distal end was fully constrained in both translation and rotation. The FE analysis, performed in Abaqus/Standard, provided spatial distributions of interstitial fluid velocity and octahedral shear strain, which were subsequently transferred to the agent-based model to regulate cell activity and tissue differentiation.

*Table 1. Material properties assigned to tissues, scaffold, and fixation components.*

| Property | Callus (granulation) | Fibrous tissue | Cartilage | Immature bone | Mature bone | Cortical bone |
|---|---|---|---|---|---|---|
| Young's modulus (MPa) | 0.2 | 2 | 10 | 1,000 | 17,000 | 17,000 |
| Permeability ($10^{-14}$ m⁴/Ns) | 1 | 1 | 0.5 | 10 | 37 | 0.001 |
| Poisson's ratio | 0.167 | 0.167 | 0.3 | 0.3 | 0.3 | 0.3 |
| Bulk modulus of grains (MPa) | 2,300 | 2,300 | 3,700 | 13,940 | 13,940 | 13,920 |
| Bulk modulus of fluid (MPa) | 2,300 | 2,300 | 2,300 | 2,300 | 2,300 | 2,300 |

### 2.2.2. Agent-based model

We employed a previously established mechanobiological bone regeneration model that has been shown to reproduce scaffold-guided bone healing in various experimental settings [23, 24, 26]. This model is a three-dimensional agent-based framework, implemented in C++ and coupled with the finite element model described in Section 2.2.1. Individual agents, each representing a 100 μm voxel, could correspond to a single cell phenotype, progenitor cell, fibroblast, chondrocyte, or immature/mature osteoblast, together with their respective extracellular matrix types: granulation tissue, fibrous tissue, cartilage, and immature or mature bone. At each daily time step, cells could proliferate, migrate, differentiate, or undergo apoptosis according to experimentally derived rates (Table 2) [32].

*Table 2. Cell activity parameters implemented in the agent-based model.*

| Cell type | Proliferation rate (day⁻¹) | Apoptosis rate (day⁻¹) | Differentiation rate (day⁻¹) | Migration speed (μm/h) |
|---|---|---|---|---|
| **Progenitor cells** | 0.6 | 0.05 | 0.3 | 30 |
| **Fibroblasts** | 0.55 | 0.05 | – | – |
| **Chondrocytes** | 0.2 | 0.1 | – | – |
| **Osteoblasts** | 0.3 | 0.16 | – | – |

Cell differentiation was regulated by the local mechanical stimulus $S$, defined as a weighted combination of octahedral shear strain $\gamma$ and interstitial fluid velocity $v$ calculated in the FE model:

$$S = \frac{\gamma}{a} + \frac{v}{b} \quad (1)$$

where $a = 0.0375$ and $b = 0.003$ mm/s are scaling constants [33, 34]. The resulting values determined whether progenitor cells differentiated toward fibrous tissue, cartilage, immature bone, or mature bone, or underwent resorption under very low stimulus conditions (Table 3) [24]. Differentiation events were linked to matrix deposition, thereby progressively altering the spatial distribution of tissue types within the callus and scaffold pores.

*Table 3. Mechano-regulation thresholds guiding progenitor cell differentiation.*

| Stimulus range (S) | Phenotype outcome |
|---|---|
| $S \leq 0.01$ | Bone resorption |
| $0.01 < S \leq 0.53$ | Mature bone |
| $0.53 < S \leq 1$ | Immature bone |
| $1 < S \leq 3$ | Cartilage |
| $S > 3$ | Fibrous tissue |

### 2.2.3 Coupling strategy

The FE and ABM components were linked through a dynamic, two-way iterative workflow (Figure 3b). At the start of each iteration, poroelastic FE analysis provided spatial maps of shear strain and fluid velocity under physiological loading. These values were imported into the ABM to guide cellular processes, including proliferation, migration, differentiation, and apoptosis. The updated tissue distribution modified the local mechanical properties, which were recalculated using a rule-of-mixtures approach and then passed back to the FE model. The new mechanical environment was resolved, and the cycle repeated. One iteration corresponded to one day of healing, allowing the coupled framework to capture the spatiotemporal evolution of scaffold-guided bone repair.

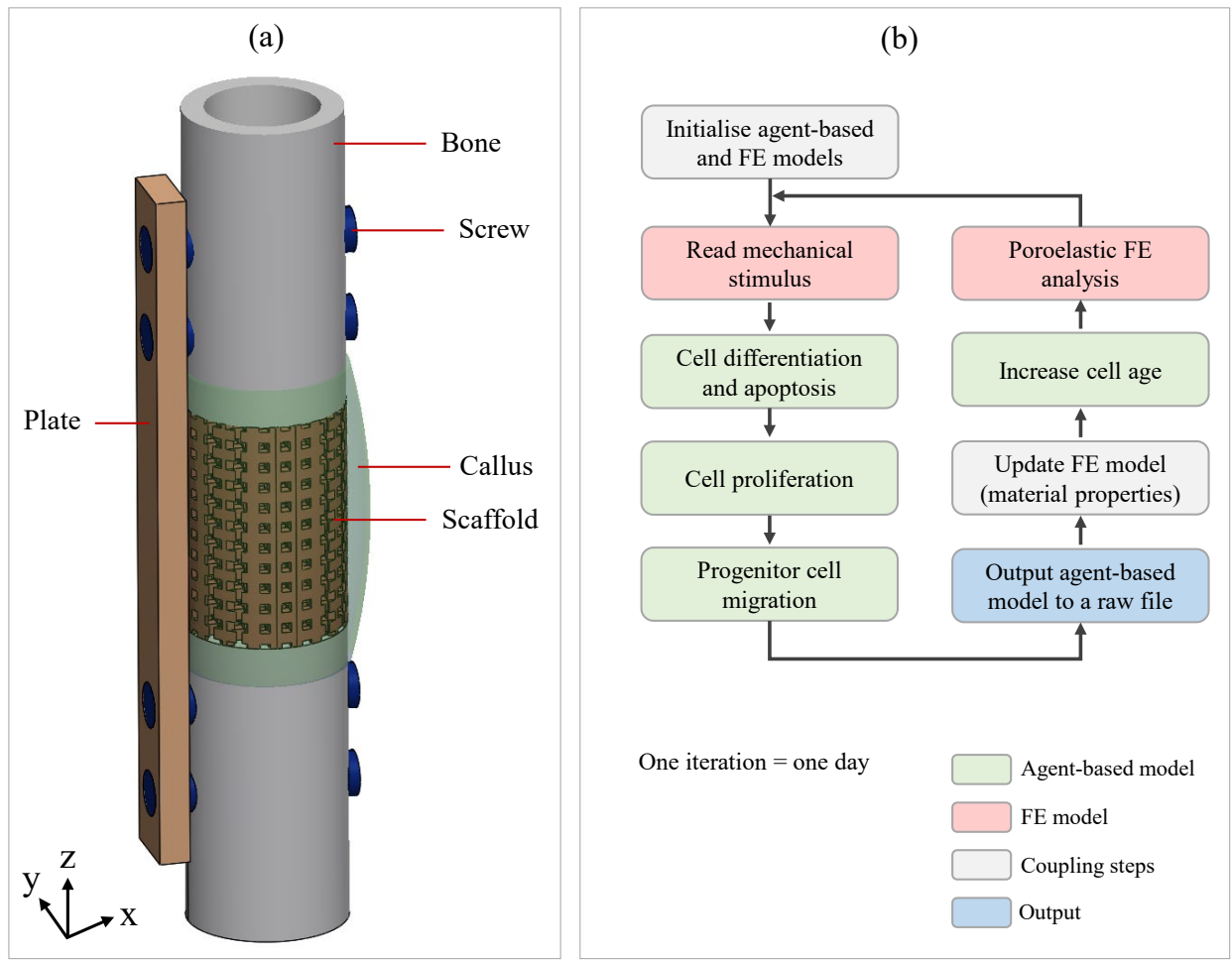

*Figure 3. Integration of finite element (FE) and agent-based modelling for scaffold-guided bone regeneration. (a) Schematic of a scaffold implanted within a bone defect stabilised by a fixation plate. (b) Coupled computational workflow: mechanical stimuli are obtained from poroelastic FE analysis and transferred to the agent-based model, where cell differentiation, apoptosis, proliferation, and progenitor migration are simulated. The updated cell distribution alters scaffold material properties, which are fed back into the FE model in an iterative loop until regeneration outcomes are predicted.*

## 3. Results and Discussion

Successful repair of critical-sized bone defects relies on scaffolds that can effectively guide and sustain bone regeneration within the defect site. Despite extensive progress in scaffold design, predicting how architectural gradients influence the spatiotemporal progression of new tissue formation remains challenging. To address this, we applied our validated FEA–ABM framework [26] to simulate the dynamics of cellular activity and tissue formation within functionally graded scaffolds.

Figure 4 illustrates the progression of regeneration within axial gradient scaffolds over 150 days at five representative time points, shown as cross-sections through the scaffold mid-plane in the x–z plane. The top row corresponds to the uniform scaffold, while the middle and bottom rows show axial gradient designs in which pore size increases (Type 1) or decreases (Type 2) toward the central region. Scaffold struts and intact host bone are shown in black and dark gray, respectively. Cell populations are represented by distinct colours: green indicates mesenchymal stromal cells (MSCs), yellow mature osteoblasts, light grey immature osteoblasts, red chondrocytes, and blue fibroblasts, while white denotes unfilled space. Together, these maps provide a visual overview of how different axial pore-size gradients influence the timing and distribution of cellular activities during bone healing.

Up to Day 30, bone formation remained minimal across all designs, with MSCs predominantly occupying the peripheral scaffold regions near the host bone. By Day 60, the first row of pores adjacent to the host bone at both the proximal and distal ends became populated with mature osteoblasts, marking the onset of bone deposition. As healing advanced, bone ingrowth progressed steadily from the defect boundaries toward the scaffold centre. By Day 150, nearly the first three pore rows were completely filled with mature osteoblasts, while the fourth row contained a mixture of MSCs and newly differentiated bone cells. Although the overall temporal pattern of bone formation was consistent across the three scaffold types, quantitative differences were evident. Type 2 scaffolds, characterised by larger pores at the host bone interface, exhibited greater filling of the initial pore rows compared with both the uniform and Type 1 designs, leading to a higher net volume of mature bone formation.

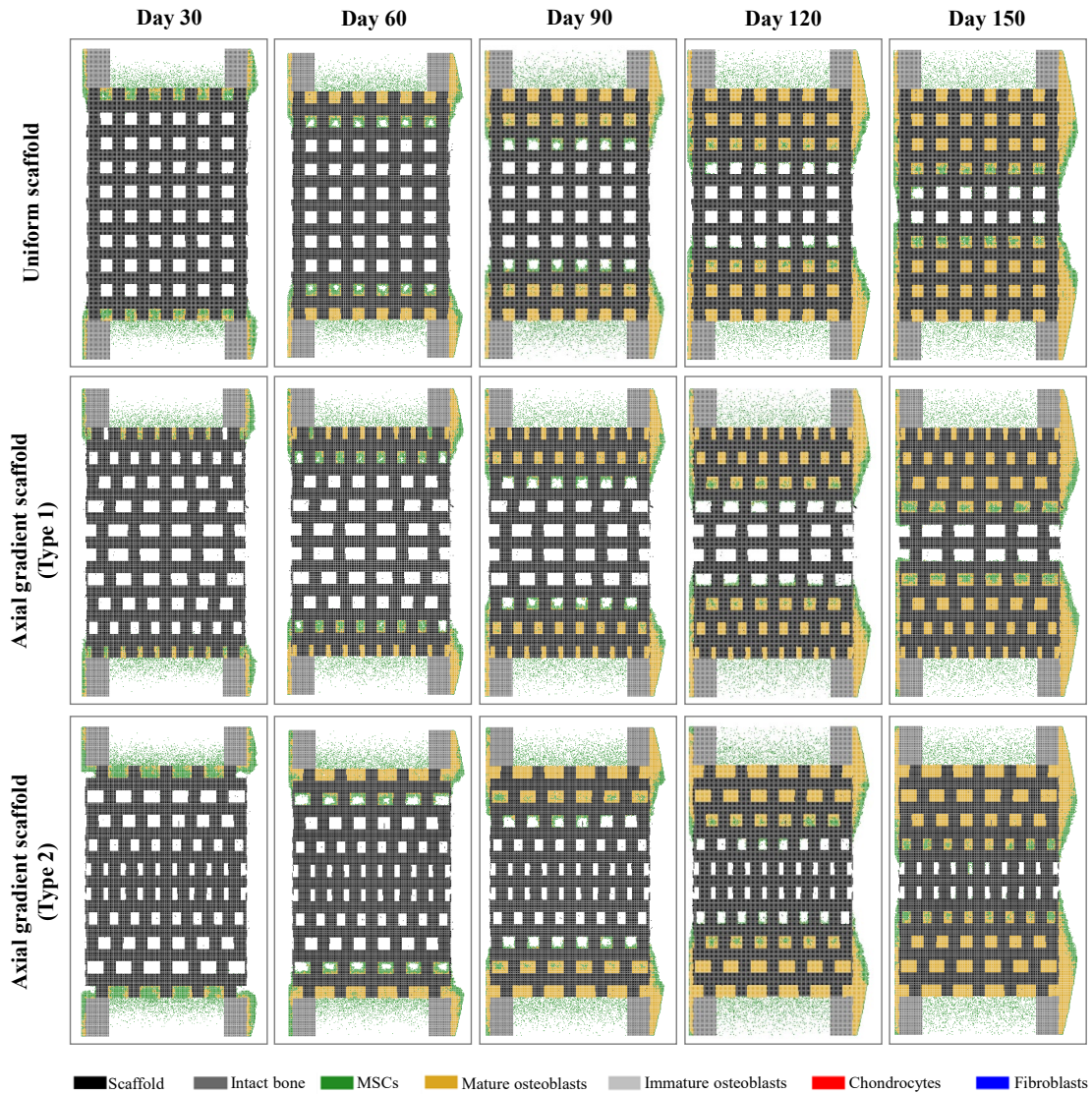

*Figure 4. Temporal progression of tissue regeneration within axial gradient scaffolds, simulated using the coupled agent-based and finite element framework. Each row corresponds to a different axial gradient scaffold type, and columns represent sequential time points during healing. The results illustrate how axial pore-size gradients influence spatial and temporal patterns of tissue ingrowth and remodelling. The results are shown as cross-sections through the scaffold mid-plane in the x–z plane. Colours are: scaffold (black), intact bone (dark grey), mesenchymal stromal cells—MSCs (green), mature osteoblasts (yellow), immature osteoblasts (light grey), chondrocytes (red), and fibroblasts (blue); white denotes unfilled pore/callus space.*

In contrast to the axial gradient designs, Figure 5 highlights regeneration within radial gradient scaffolds, where pore-size gradients were oriented from the periphery toward the centre: Type 1 with larger central pores and Type 2 with smaller central pores, alongside the uniform reference scaffold. The results are shown as cross-sections through the scaffold mid-plane in the x–z plane.

During the early phase (Day 30), bone formation was negligible, with MSCs concentrated along the cortical interface. By Day 60, mature osteoblasts began to populate the outer pore layers, marking the onset of bone deposition from the periphery inward. From Day 90 onward, bone ingrowth progressed toward the scaffold centre, though the depth of penetration varied between designs.

By Day 150, both the uniform and Type 1 scaffolds exhibited gradual inward progression, with bone formation more pronounced at the sides, reflecting the influence of peripheral deposition. In contrast, Type 2 showed faster central ingrowth than the other two designs. Its smaller central pores likely offered greater surface area for cell attachment and bone growth, facilitating deeper infiltration toward the scaffold core [35]. Despite these spatial differences, the total amount of bone formed remained broadly comparable across all three scaffold types.

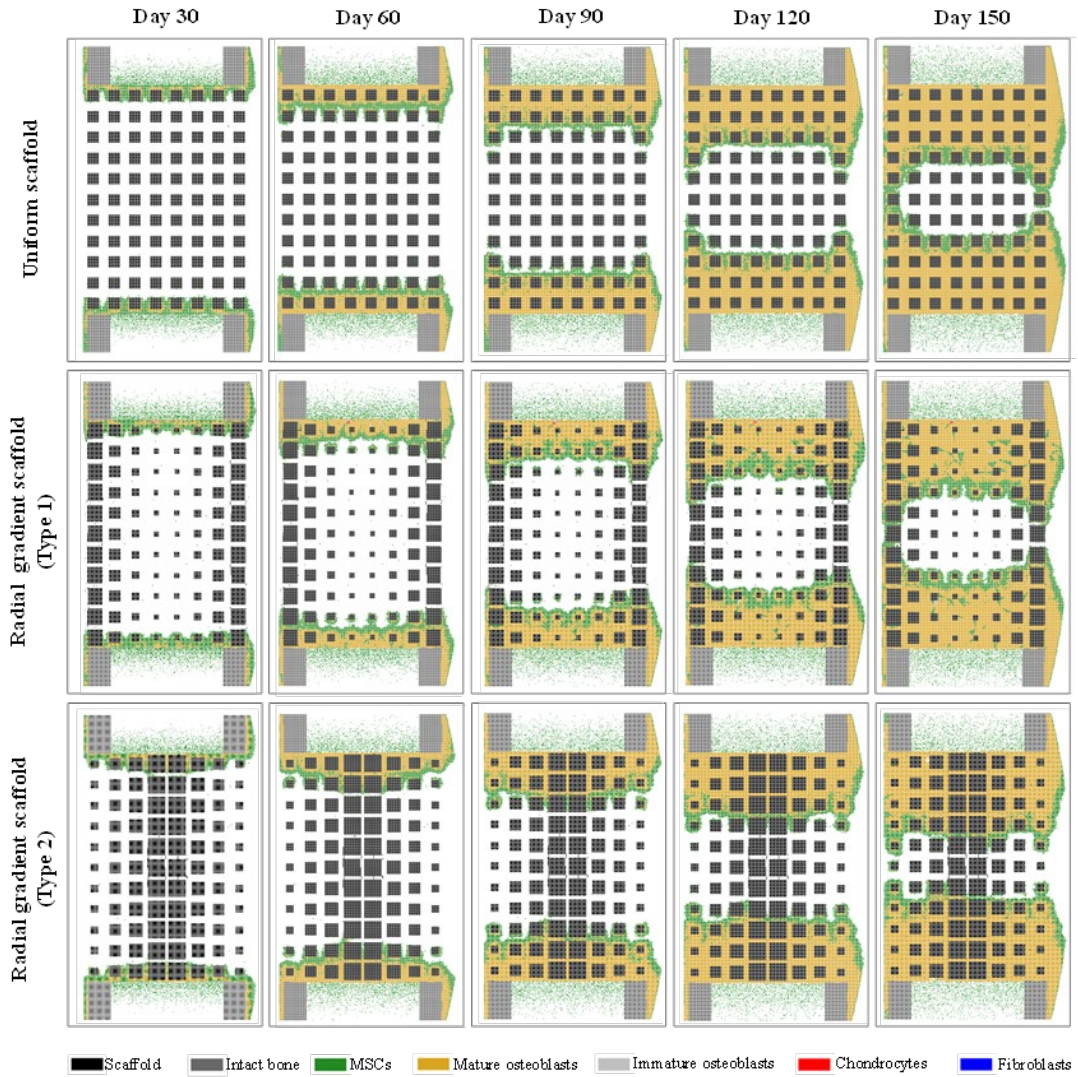

*Figure 5. Temporal progression of tissue regeneration within radial gradient scaffolds, simulated using the coupled agent-based and finite element framework. Each row corresponds to a different radial gradient scaffold type, and columns represent sequential time points during healing. The results illustrate how axial pore-size gradients influence spatial and temporal patterns of tissue ingrowth and remodeling. The results are shown as cross-sections through the scaffold mid-plane in the x–z plane. Colors are: scaffold (black), intact bone (dark gray), mesenchymal stromal cells—MSCs (green), mature osteoblasts (yellow-ochre), immature osteoblasts (light gray), chondrocytes (red), and fibroblasts (blue); white denotes unfilled pore/callus space.*

To complement the cross-sectional maps (Figures 4–5), Figure 6 reports quantitative bone ingrowth as a percentage of scaffold pore volume over time. In the axial gradient designs (Figure 6a), all scaffolds showed a monotonic increase from ~3–4% at Day 30 to 60–76% at Day 150. Despite having similar overall porosity, Type 2 yielded the highest occupancy at every time point (≈21%, 39%, 59%, and 77% at Days 60, 90, 120, and 150, respectively), the uniform scaffold was intermediate (≈17%, 35%, 52%, 69%), and Type 1 remained lowest (≈14%, 28%, 43%, 60%). These trends are consistent with the cross-sectional observations: although the depth of ingrowth was comparable across designs, the larger pores positioned near the proximal and distal bone interfaces in Type 2 facilitated greater overall bone deposition, whereas the smaller pores in Type 1 limited net bone infill.

For the radial designs (Figure 6b), ingrowth also increased steadily over time. However, differences in net bone formation were less pronounced, particularly between the uniform scaffold and Type 2, which exhibited nearly identical values at later stages. Type 1 performed comparably to the other two designs at early time points, but by Day 150 its occupancy lagged slightly behind. Taken together, these results demonstrate that gradient scaffolds, and the orientation of the gradient (axial versus radial), can significantly influence the amount of bone formation.

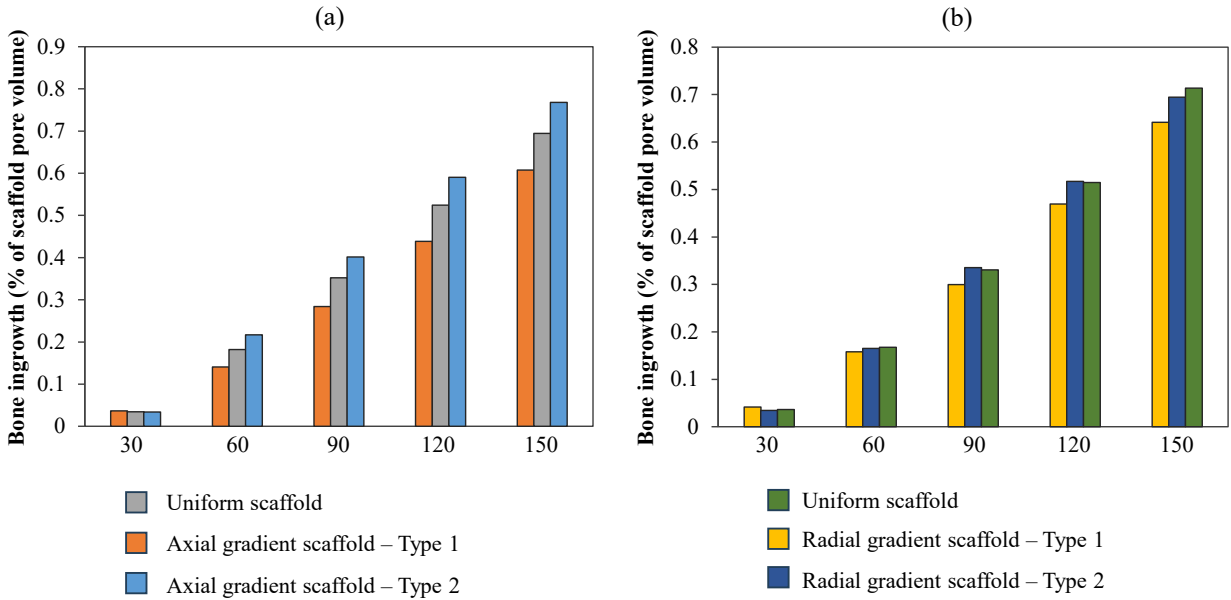

*Figure 6. Quantitative evaluation of tissue regeneration within scaffolds over time. (a) Comparison of tissue formation in uniform scaffolds, axial gradient scaffolds (Type 1), and axial gradient scaffolds (Type 2). (b) Comparison of tissue formation in uniform scaffolds, radial gradient scaffolds (Type 1), and radial gradient scaffolds (Type 2). Data are shown as regenerated tissue volume normalized to the void space within the scaffolds across sequential time points, illustrating how scaffold architecture influences tissue ingrowth dynamics.*

In addition to supporting bone regeneration, scaffolds must also provide sufficient mechanical stability to withstand physiological loading [36]. Therefore, it is essential to evaluate both their regenerative capacity and structural performance. To this end, we conducted FEA to assess the von Mises stress distribution within the different scaffold architectures under loading conditions.

Figure 7 presents the von Mises stress maps for the uniform scaffold and the two axial gradient designs. The top row shows cross-sections through the scaffold mid-plane in the x–z plane, while the bottom row shows cross-sections in the y–z plane. Stresses were consistently higher on the lateral side (positive x-direction) of the x–z plane, since the opposite side was constrained by the fixation plate while loading was transmitted through the free side. In the y–z plane, elevated stresses were observed on the anterior side (positive y-direction) due to the applied bending moment. Stress concentrations were most pronounced at the anterior–lateral quadrant, localised around strut junctions, while the majority of struts experienced relatively low stresses.

Compared with the uniform design, the two gradient scaffolds redistributed stresses more heterogeneously along the scaffold height. Type 2 exhibited elevated stresses near the proximal and distal ends, where pores were larger, whereas Type 1 displayed higher stresses near the mid-height region, reflecting the larger pores present in this area. Overall, these findings indicate that axial pore-size gradients not only influence tissue regeneration but also modify how mechanical loads are transferred through the scaffold.

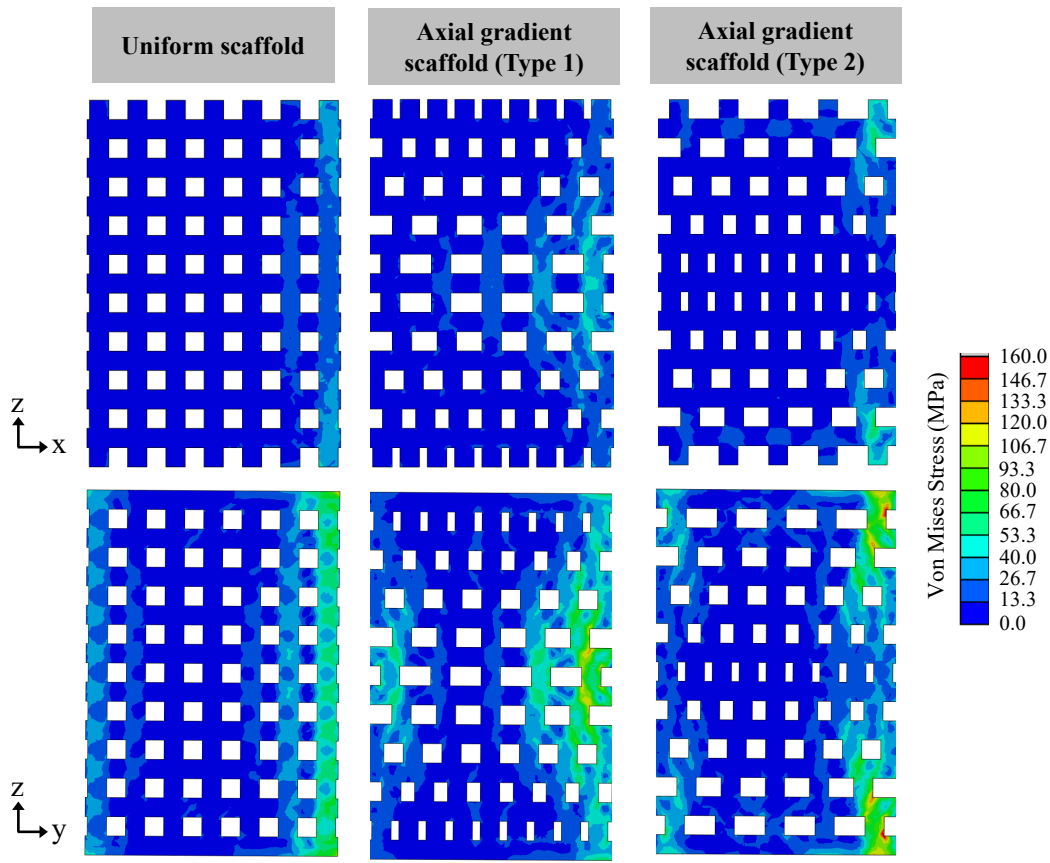

*Figure 7. Von Mises stress distribution within uniform and axial gradient scaffolds (Type 1 and Type 2) under loading.* The top row shows the stress distribution across the mid-cross-section in the x–z plane, while the bottom row presents the distribution across the y–z plane, demonstrating the influence of scaffold architecture on stress distribution.

Figure 8 shows the von Mises stress distributions for the uniform scaffold and the two radial gradient designs. As in the axial case, stresses concentrated on the lateral side (positive x-direction) in the x–z plane and the anterior side (positive y-direction) in the y–z plane, consistent with the load path and bending moment.

Radial gradients strongly modulated the peripheral stress distribution. Type 1, with smaller pores and denser struts at the scaffold boundary, exhibited the lowest stresses overall. The stiffer outer ring increased local bending stiffness and effective load-bearing area, distributing stresses more evenly and reducing peak values. In contrast, Type 2, which had larger peripheral pores and thinner struts at the boundary, showed higher localised stresses at the periphery, particularly in the anterior–lateral quadrant. The uniform scaffold displayed an intermediate pattern between these two extremes.

Together, these results demonstrate that while axial gradients (Figure 7) primarily shifted stress accumulation along the scaffold height, radial gradients (Figure 8) determined the magnitude of stresses at the periphery. A denser boundary network, as in Type 1, buffered applied loads and minimised peak stresses, whereas a more open periphery, as in Type 2, increased them.

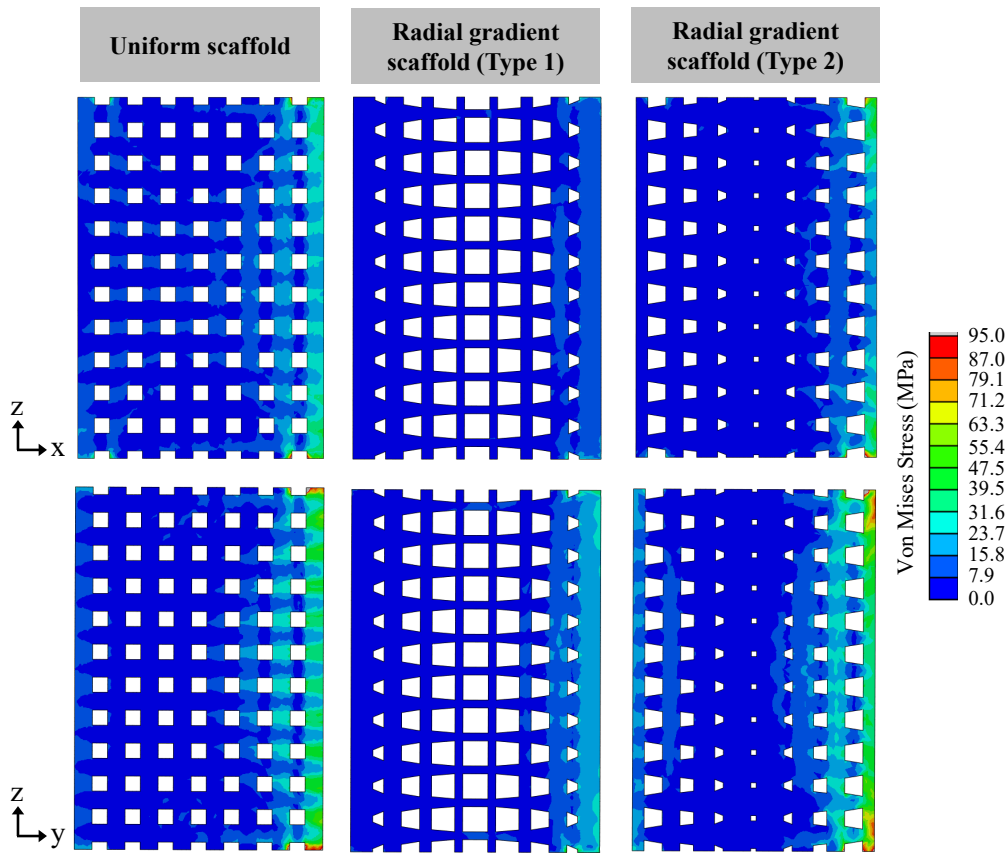

*Figure 8. Von Mises stress distribution within uniform and radial gradient scaffolds (Type 1 and Type 2) under loading.* *The top row shows the stress distribution across the mid-cross-section in the x–z plane, while the bottom row presents the distribution across the y–z plane, demonstrating the influence of scaffold architecture on stress distribution.*

Figure 9 summarises the peak von Mises stresses for all scaffold types, complementing the stress distribution maps in Figures 7 and 8. For the axial gradients (Figure 9a), both Type 1 and Type 2 exhibited higher peak stresses than the uniform scaffold. Type 2 reached the highest value (~160 MPa), reflecting the stress accumulation at the proximal and distal ends where pores were larger, while Type 2 showed a slightly lower peak (~150 MPa) due to stress concentration at the mid-height region. The uniform scaffold, with a more homogeneous structure, maintained the lowest peak stress (~120 MPa).

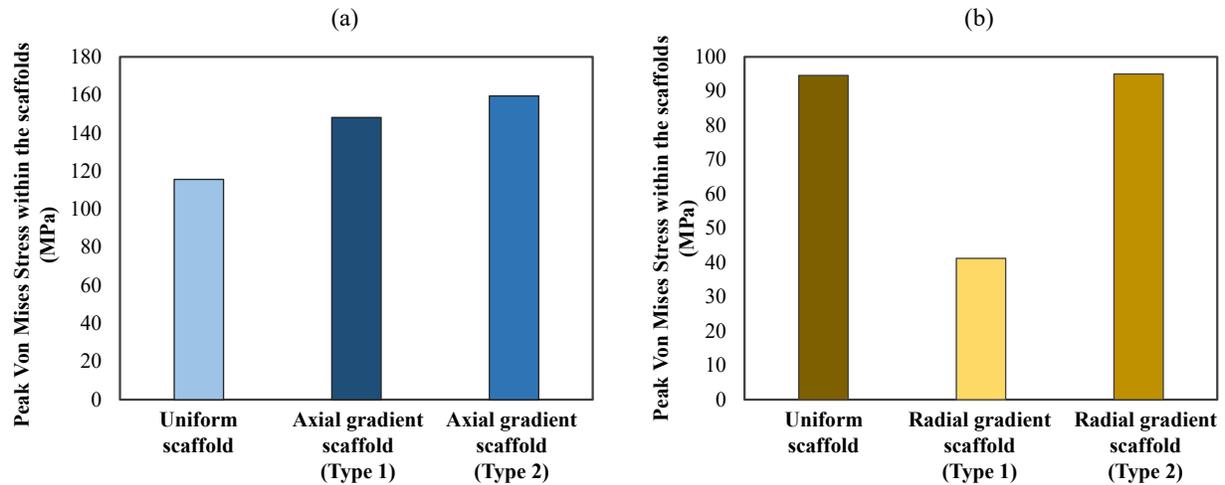

*Figure 9. Peak von Mises stress in scaffolds with different architectures.* (a) Axial gradient scaffolds compared with the uniform scaffold. (b) Radial gradient scaffolds compared with the uniform scaffold.

For the radial gradients (Figure 9b), a contrasting trend was observed. Type 1, with smaller pores at the periphery, showed the lowest peak stress (~40 MPa), confirming that a denser outer ring reduces local stress magnitudes by providing greater stiffness and load-sharing capacity. In contrast, Type 2, with larger peripheral pores, exhibited elevated peak stresses (~95 MPa), comparable to the uniform scaffold (~93 MPa).

Together, these results highlight a key distinction: axial gradients altered where along the height stresses accumulated, increasing peak values, while radial gradients primarily modulated peripheral stiffness, with a denser outer boundary (Type 1) markedly reducing peak stresses. These findings emphasise the importance of not only incorporating gradients but also carefully selecting their orientation and location to balance mechanical stability with regenerative performance.

A central outcome of this study was the identification of a fundamental design trade-off: axial gradients enhanced regenerative capacity by accelerating bone ingrowth, particularly when larger pores were placed at the bone–scaffold interface (Type 2), whereas radial gradients improved structural competence by reducing stress concentrations through a denser peripheral ring (Type 1). This finding highlights that scaffold design cannot prioritise mechanics or biology in isolation, but must instead adopt a multi-objective optimisation strategy that balances both requirements.

Mechanistically, these patterns arise from the interplay between local pore architecture, load transfer, and cell–matrix interactions. In the axial gradients, larger pores positioned at the host bone interface provided additional space for cellular infiltration and matrix deposition, thereby facilitating greater net bone growth. By contrast, the pore-size distribution in radial gradients did not confer a similar advantage for tissue ingrowth, aligning with recent in vivo findings showing that bone formation in radially graded scaffolds did not outperform uniform controls [35].

Our results extended prior work by advancing hybrid ABM–FEA approaches from uniform scaffolds to functionally graded architectures evaluated under physiologically relevant loading [20]. While in vitro studies have demonstrated that pore-size gradients enhance osteogenic differentiation and cell adhesion [14, 37], they have not simultaneously addressed the mechanical implications within in vivo-like environments. By explicitly integrating poroelastic FEA with ABM, our framework not only captured how architectural gradients influence cellular dynamics but also quantified the resulting structural consequences. This mechanistic platform may complement emerging machine learning–driven optimisation strategies by providing biologically and mechanically grounded datasets that can inform, train, and validate data-driven models [22, 38].

From a translational standpoint, different gradient strategies may be optimal depending on the clinical context. For load-bearing defects such as long bones, radial gradients with stiffer peripheries may be more suitable to ensure mechanical stability. Conversely, defects in metaphyseal or cancellous regions, or in clinical situations where rapid bridging is critical, may benefit from axial gradients with larger pores at the host interface to maximise bone regeneration. Ultimately, tailoring scaffold design to defect location, patient anatomy, and load environment will be key to clinical translation.

This study also has limitations. The scaffold material was assumed to be titanium, which does not account for degradability or remodelling as occurs with bio ceramics or polymers [39]. Biological rules for cell differentiation were derived from generalised thresholds rather than patient-specific data, and vascularisation was not explicitly modelled despite its critical role in regeneration. Moreover, while our framework provides mechanistic predictions, future animal studies are needed to validate these computational observations and confirm their translational relevance. Future work should address these limitations by extending the framework to degradable materials, incorporating patient-specific geometries and loading conditions, and coupling with angiogenesis models. In addition, integration with optimisation algorithms could enable automated discovery of scaffold designs that simultaneously maximise mechanical and biological outcomes.

## 4. Conclusion

This work establishes an integrated finite element–agent-based modeling (FEA–ABM) framework to simultaneously interrogate the mechanical and biological performance of functionally graded (FG) bone scaffolds under physiologically relevant conditions. By unifying poroelastic stress analysis with explicit simulation of progenitor cell dynamics, the framework captures both load transfer pathways and the spatiotemporal progression of bone formation, thereby providing a mechanistic view of scaffold-guided regeneration.

The simulations revealed that gradient orientation exerts a decisive influence on scaffold function. Axial gradients with enlarged pores at the bone–scaffold interface (Type 2) accelerated bone ingrowth and enhanced regenerative capacity, whereas radial gradients with denser peripheral struts (Type 1) mitigated stress concentrations and improved mechanical stability. These findings underscore a fundamental design trade-off between promoting tissue regeneration and maintaining structural competence, emphasising that gradient strategies must be tailored to the clinical context and defect environment.

By integrating cell-scale biological processes with macroscale mechanical analysis, this framework advances beyond conventional FEA- or mechanobiology-only models, offering a rational pathway for the optimisation of scaffold architectures prior to experimental validation. Future extensions incorporating patient-specific geometries, manufacturing constraints, and in vivo data will further strengthen its translational potential, supporting the development of next-generation FG scaffolds for reliable repair of large bone defects.